\documentclass[letterpaper]{article}
\usepackage{opex3}
\begin{document}

%%%%%%%%%%%%%%%%%% title page information %%%%%%%%%%%%%%%%%%
\title{Photophysics of novel diamond based single photon emitters}

%%%%%%%%%%%%%%%%%%%%%%%%%%%%%%%%%%%%%%%%%
\author{I. Aharonovich, S. Castelletto, D. A. Simpson, A. D. Greentree and \\ S. Prawer}
\address{School of Physics, The University of Melbourne, 3010
Victoria, Australia}
\email{i.aharonovich@pgrad.unimelb.edu.au, sacas@unimelb.edu.au} %% email address is required
%%%%%%%%%%%%%%%%%%%%%%%%%%%%%%%%%%%%%%%%%


%%%%%%%%%%%%%%%%%%%%%%%%%%%%%%%%%%%%%%%%%
\begin{abstract}
A detailed study of the photophysical properties of several novel color centers in chemical vapor deposition diamond is presented. These emitters show narrow luminescence lines in the near infra-red. Single photon emission was verified with continuous and pulsed excitation with emission rates at saturation in the MHz regime, whilst direct lifetime measurements reveal excited state lifetimes ranging from 1-14 ns.
In addition, a number of quantum emitters demonstrate two level behavior with no bunching present in the second order correlation function. An improved method of evaluating the quantum efficiency through the direct measurement of the collection efficiency from two level emitters is presented and discussed. \\
\end{abstract}
%%%%%%%%%%%%%%%%%%%%%%%%%%%%%%%%%%%%%%%%%

\ocis{(270.0270) Quantum optics; (270.5290) Photon Statistics;
(160.4760) Optical properties; (170.1790) Confocal microscopy}
\hfill
% REPLACE WITH CORRECT OCIS CODES FOR YOUR ARTICLE

%%%%%%%%%%%%%%%%%%%%%%% References %%%%%%%%%%%%%%%%%%%%%%%%%
%%%%%%%%%%%%%%%%%%%%%%% References %%%%%%%%%%%%%%%%%%%%%%%%%

%%%%%%%%%%%%%%%%%%%%%%%%%%%%%%%%

\section{Introduction}
The emerging field of quantum optics has established a demand for an
accessible solid state system which can generate a stream of single
photons on demand \cite{Bouwmeester07}. Although single photon
emission has been demonstrated from quantum dots (QD)
\cite{Bouwmeester07,Santori01},single molecules \cite{Brunel99} and
nanowires\cite{Tribu08}, the operation of those systems is often
limited by the temperature. Diamond crystals, on the other hand, offer the most
promising platform for generation of robust, photo stable, single
photons at room temperature \cite{Kurtsiefer00,Beveratos02b}.
Nevertheless, out of more than 500 existing optical centers in
diamond \cite{Zaitsevhandbook}, only three known centers
demonstrate single photon emission, namely, the nitrogen-vacancy
complex \cite{Jelezko06,Beveratos02b}, the silicon-vacancy complex
\cite{Wang06} and the nickel-nitrogen complex
\cite{Gaebel04,Wu06,Wu07}.

Recent progress in materials
science and fabrication techniques of diamond color centers \cite{Aharonovich08} unveiled
novel quantum emitters operating in the MHz regime
\cite{Simpson09,Aharonovich09}. These new centers, are comparable to QD in
terms of brightness \cite{Bouwmeester07}, but have the
tremendous advantage of stable operation at room temperature.
These break throughts have opened a new avenue to investigate essentially an unknown group of single photon emitters in diamond.

The simple two-level nature of a select few single photon emitters allows a direct estimate of the collection efficiency
associated with the optical setup, the precise knowledge of the collection efficiency can lead to more accurate estimations of the quantum efficiency of the more common three level single photon emitting systems.  This type of source can thus eventually be used as a reference for an absolute characterization of the quantum efficiency along with more precise estimations of the collection efficiency of the system, provided the photo-physics is well understood. One can envisage that single photon sources could be implemented as a ``single photon standard'', able to link classical radiometric measurements to the fundamental quantum optical entities. In particular in the long term such a single photon standard could contribute to a re-definition of the standard units for optical radiation in terms of the ``quantum candela'' \cite{Cheung07}.
The controlled fabrication of the MHz class of diamond single photon emitters, together
with a recent emission enhancement by coupling the light to plasmonic structures \cite{Schietinger09}
or cavities \cite{benson08}, is expected
to progress the technology beyond the point in which they can become practical quantum information and metrology devices.

In this paper, a detailed study of the photophysical properties of the novel family of single photon emitters reported
 in \cite{Aharonovich09} has been undertaken. To gain insight into the electronic level structure of the emitters,
 their fundamental optical properties such as lifetime, photo-luminescence (PL) spectrum, sub-poissionian statistics
 and saturation count rate have been studied and compared. We concentrate on the typical PL lines in the range of 740-770 nm,
 which exhibit single photon emission with count rates ranging from 1.3-3.2 $10^6$ counts/s at saturation.
\section{Theoretical background}
\begin{figure}[htbp]
\centering\includegraphics[width=12cm]{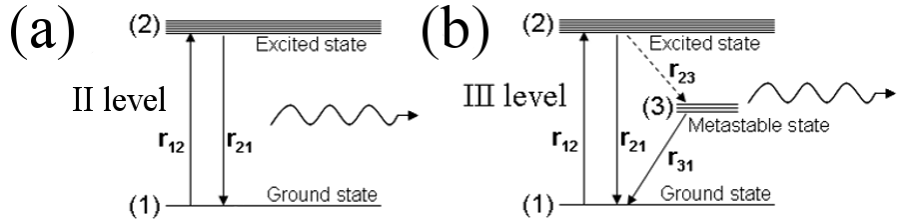} \caption{Schematic
diagram of a (a) two-level and (b) three-level system, where
$r_{nm}$ $n,m=1,2,3$ are the transition rates from level $(n)$ to
level $(m)$.}\label{Two and three level diagram}
\end{figure}

The physics of the photoluminescence emitted from a single diamond color center
can be described within the framework of a two level system comprising a ground and excited state,
and in many instances as a three-level system \cite{Kurtsiefer00,Wu06,kitson98}, whereby the single center is excited from its
ground state to the excited state, with a third longer lived state
providing an additional decay path from the excited state, as seen in Fig. \ref{Two and three level diagram}.
The rate equations describing the populations of the two and three
level systems can be written in matrix form as:
\begin{equation}
\left(
  \begin{array}{c}
    \dot{n_1} \\
   \dot{n_2} \\
  \end{array}
\right)=\left(
          \begin{array}{cc}
            -r_{12} & r_{12} \\
           r_{12} & -r_{21}\\
          \end{array}
        \right)\left(
                 \begin{array}{c}
                   n_1 \\
                   n_2 \\
                 \end{array}
               \right)
 \label{RE of a two level system}
\end{equation}

and
\begin{equation}
\left(
  \begin{array}{c}
    \dot{n_1} \\
    \dot{n_2} \\
    \dot{n_3} \\
  \end{array}\right)=\left(
                \begin{array}{ccc}
                  -r_{12}&r_{21}& r_{31} \\
                 r_{12}&-r_{21}-r_{23}&0\\
                  0&r_{23}&-r_{31}\\
                \end{array}
              \right)\left(
                       \begin{array}{c}
                         n_1 \\
                         n_2\\
                        n_3 \\
                       \end{array}
                     \right).
\label{RE of a three level system}
\end{equation}

By solving Eqs (\ref{RE of a two level system}) and (\ref{RE of a
three level system}) with the initial condition $n_1=1, n_2=0,
n_3=0$ (i.e the system is prepared in the ground state), the
instantaneous emission probability of a photon being emitted from
the excited state, $n_2(t)$, can be obtained. By normalizing
$n_2(t)$ to the probability of a photon being emitted at an infinite
time $n_2(\infty)$ an analytical expression describing the second
order correlation function can be obtained. The respective
analytical expressions for the two and three level
\cite{Wu06,kitson98} cases are then given by:
\begin{equation}
g^{(2)}(\tau)= 1- \exp(- \lambda_1 \tau). \label{g2 eq1}
\end{equation}
where $\lambda_1= r_{12}+r_{21}$ and
\begin{equation}
g^{(2)}(\tau)= 1- (1+a)~ \exp(- \lambda_1 \tau)+ a ~
\exp(- \lambda_2 \tau), \label{g2 eq2}
\end{equation}
where $\lambda_2= r_{31}+r_{23} r_{12}/\lambda_1 $, and $a=r_{12} r_{23}/(\lambda_1 r_{31})$. \\\\
The decay rate $\lambda_1$ depends
on the excitation optical power, $P_{\mathrm{opt}}$ as:
\begin{equation}
\lambda_1=r_{21}^0 (1 + \alpha P_{\mathrm{opt}}), \label{eq4}
\end{equation}
where $r_{21}^0$ is the inverse of the excited state lifetime ($\tau_{21}$) and $\alpha$ is a fitting parameter.\\
\\
Similarly, for the three-level system, the decay rate $\lambda_2$, assuming
$r_{23}$ constant with the excitation power \cite{Wu06}, can be written in terms of $r_{31}^0$ in the
limit of zero optical power.
\begin{equation}\label{eq4b}
\lambda_2=r_{31}^0(1+\beta P_{\mathrm{opt}})+ \frac{r_{23}
r_{12}}{r_{21}^0(1+ \alpha P_{\mathrm{opt}})},
\end{equation}
where $\beta$ is a fitting parameter.

 With an accurate determination of the second order
correlation function one can determine the individual decay rates
involved in the system and gain an understanding of the
energy level structure. Upon inspection of the two and
three level auto-correlation functions there is a clear distinction
in the shape and nature of the exponential component. In the two
level case the function exhibits a simple exponential rise at a rate
equivalent to the fluorescence decay rate of the excited state in
the limit of zero optical excitation i.e. $r_{12}\rightarrow0$, with
the exponential asymptotic to a g$^{(2)}$($\tau$) value of 1. The
three level expression, on the other hand, contains two exponential
components each with a characteristic time constant. Depending on
the transition rates from the excited state to the shelving state
and shelving state back to the ground state, g$^{(2)}$($\tau$) can
increase beyond 1 for times $> \tau_{21}$, before decaying
back to 1 at times $>> \tau_{21}$.

The phenomenon of
g$^{(2)}$($\tau$)>1 is commonly termed ``photon bunching" and this
behavior enables a clear distinction to be made between two and
three level systems. The second order correlation function describes
the probability of detecting a photon with a delay time $\tau$ after one photon has been detected at $\tau$=0.
Hence, the bunching effect describes the enhanced probability to detect a photon at short times then at longer times. Indeed after the
system undergoes a transition to its shelving state, which has longer lifetime, there is a longer interval between the photons. Once the system
relaxed to the ground state and undergo a full emission cycle again, the normal photon rate is achieved again. The waiting interval while the system in the metastable state thus creates the bunching effect of the g$^{(2)}$($\tau$).

The measurement g$^{(2)}$($\tau$) is done
typically using a Hanbury-Brown and Twiss (HBT) interferometer whereby the
coincidence events of two detectors are measured. In practice, the
temporal jitter of the detectors and of the electronics can be
ignored provided that the lifetime is much longer than the temporal
jitter; with the identification of emitting centers with increasing
short fluorescence lifetimes \cite{Wang06,Wu06,Aharonovich08}, the
effect of the temporal jitter on the measured auto-correlation
function $g^{(2)}_{\mathrm{meas}}(\tau)$ must be taken into
account. Therefore the measured auto-correlation function is given
by the convolution of Eqs (\ref{g2 eq1},\ref{g2 eq2}) with the
instrumental time response function $J(\tau)$:
\begin{equation}
g^{(2)}_{\mathrm{meas}}(\tau)=\int_{-\infty}^{\infty}
\mathrm{d}\tau' g^{(2)}(\tau') J(\tau-\tau'). \label{g2 con}
\end{equation}\\
The majority of single photon emitters identified in diamond (such as: Si-V \cite{Wang06}, NV \cite{beveratos01} and the
NE8\cite{Gaebel04,Wu06}) have been attributed to three level systems
which suffer from quenching due to the presence of a shelving state. However, the recent identification of
highly efficient two level systems in diamond \cite{Aharonovich08,Simpson09} has attracted considerable interest, particularly due to its central role in determining more accurately the collection efficiency of the setup and the quantum efficiency of three level systems. The quantum efficiency, $\eta_{QE}$, of an emitter is defined as
 the probability of an absorbed pump photon resulting in emitted photon. For a two level emitter, the quantum efficiency is equal to 1 when $r_{12}$ $>>$ $r_{21}$, indicating that for each absorbed pump photon a photon is emitted from the excited state.

We can deduce the number of emitting photons or the fluorescence count rate of a two level system by solving
the rate equations in Eq. (\ref{RE of a two level system}):
\begin{equation}\label{eq3}
\phi=\frac{\phi_{\mathrm{\infty}}
P_{\mathrm{opt}}}{P_\mathrm{sat}+P_\mathrm{opt}},
\end{equation}
where $\phi$ represents the single photon count rate, $\phi_{\mathrm{\infty}}=
r_{21}^0 \eta$, is the saturation count rate for
$P_{\mathrm{opt}}\rightarrow \infty$ where $\eta$ represents the total collection
efficiency and $P_\mathrm{sat}=\alpha^{-1}$ is the optical saturation
power.

In the three level case the  quantum efficiency can be derived analytically
in terms of the detected fluorescence rate by solving the rate equations in Eq. (\ref{RE of a three level system}):
\begin{equation}
 \phi=\eta_{QE}\times \eta  \frac{r_{21}}{1+
\frac{r_{21}}{r_{12}}+ \frac{r_{23}}{r_{31}}}. \label{QE}
\end{equation}

\section{Experiment}
The CVD diamond crystals employed in this work were grown to an
average size of few hundreds nanometers from diamond seeds (4-6 nm) on a sapphire substrate using a
microwave plasma enhanced CVD technique (900 W, 150
Torr)\cite{Stacey09}. A home built confocal microscope with a
spatial resolution $\sim$ 400 nm and HBT
interferometer were used to identify the emitting centers and
measure the time correlation of photoluminescence (PL) intensity
(Fig. 2(a)). A fiber coupled CW diode laser emitting at 682 nm and
a 690 nm pulsed diode 200 ps pulse width (repetition rate from 10
MHz up to 80 MHz) were interchanged for excitation. The lasers
polarization was controlled by a Glan Taylor polarizer and a
half-wave plate. The diamond sample was mounted on a piezo XYZ stage
with 0.2 nm resolution, allowing 100x100 $\mu$m$^2$ scans. The
unwanted residual laser line was eliminated by a dichroic beam
splitter and a $F_1$ broad (10 nm FWHM) band-pass filter centred at
740 nm, 749 nm or 760 nm depending on the PL from a specific crystal. The PL from the emitting centers was then coupled into a
62.5 $\mu$m core multimode fiber, which acts as an aperture. A 50:50
fibre coupled beam splitter guided the photons to two single photon
counting detectors (APDs) and their outputs were sent to the start
and stop inputs of the time correlator card. The PL was recorded without the bandpass $F_1$ filter
using a fiber-coupled spectrometer with a cooled CCD array. All the measurements were performed at room temperature.

\begin{figure}[htbp]
\centering\includegraphics[width=11cm]{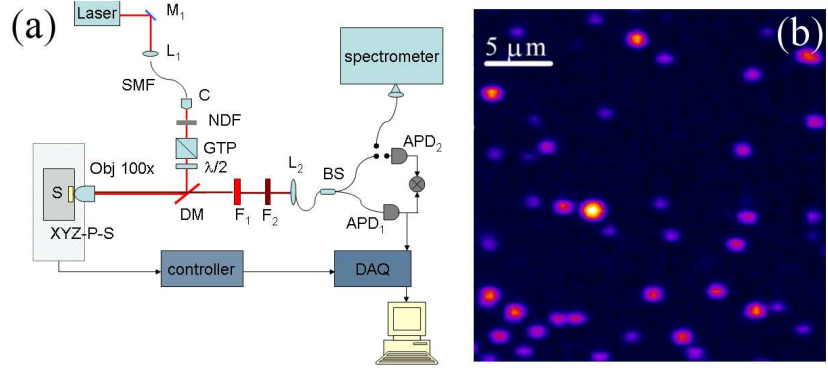}
\caption{(a)Experimental setup. A diode laser at 682 nm operating in
CW and a pulsed 690 nm laser were alternatively coupled to a single
mode fiber, and then collimated (C) and polarized by a Glan Taylor
polarizer (GTP). The laser polarization was varied by an half-wave
plate ($\lambda/2$). A variable neutral density filter (NDF) was
used to change the excitation power. Samples (S) were excited by
focusing the laser light by a high numerical aperture (0.9)
objective
 with 100x magnification. A dichroic mirror (DM)transmitting from 720 nm, was used to separate the laser line form
  the sample fluorescence emission, when collected back from the same Obj. $F_{1,2}$ are band-pass filters,
   794$\pm$80 nm and alternatively 760$\pm$12 nm or 740$\pm$12 nm, to isolate the single photon emission lines. A 100 mm focal length lens was used to send the single photon emission to a multimode-fiber, providing an aperture for the confocal imaging. Finally a 50:50 fiber beam splitter was used to verify the single photon emission by performing the auto-correlation between two low-dark counts (150 counts/s) single
photon counting modules (APD). The S were mounted on a Physics Instruments XYZ piezo stage in closed loop operation. (b) Confocal image of 20x20 $\mu$m$^2$ showing bright spots which corresponds to a color center within a diamond crystal.}
\end{figure}

\section{Results}
More than 10 different sapphire substrates were scanned, with crystals grown to a size of a few hundred nanometers.
 Fig. 2b shows a typical
confocal map of the diamond crystals obtained by 682 nm laser
excitation. The PL spectra of the bright spots revealed emission with zero phonon lines (ZPLs) centered at 744$\pm$2 nm (FWHM $\sim$ 11nm),749$\pm$2
nm (FWHM $\sim$ 4nm), 756$\pm$2 nm (FWHM $\sim$ 11 nm), and
764$\pm$2 nm (FWHM $\sim$ 10 nm), as shown in Fig. 3.  In some
cases two or more of these lines were found in one crystal,
as shown in the inset of Fig. 3. Anti-bunching measurements on
crystals containing multiple emission lines generally showed
multiple emitters with auto-correlation functions at the zero delay
time of $g^{(2)}(0)> 0.5$, with the intensity of the lines varying
from crystal to crystal. These characteristic emission lines are in
agreement with cathodo-luminescence lines from chromium related
centers in diamond, exhibiting narrow lines in the region of
740-770 nm \cite{Zaitsev00}.

\begin{figure}[htbp]
\centering\includegraphics[width=11cm]{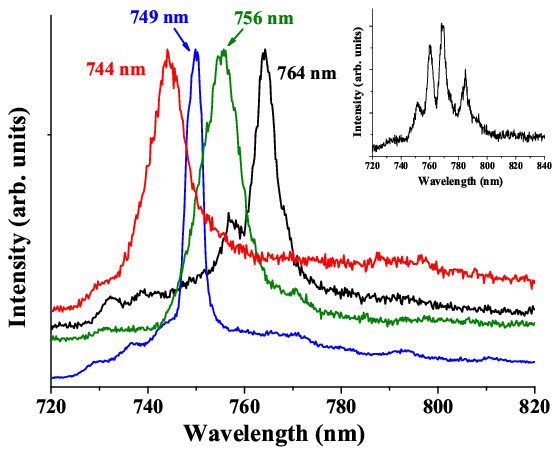} \caption{Typical PL
spectra recorded from individual CVD diamond crystals as shown in
the raster scan of the sample in Fig. 1.
 The peak emission lines centred at 744 nm (red), 749 nm (blue), 756 nm (green) and 764 mn (black).
Similar emission lines were also found all together in one crystal (inset).  }
\end{figure}

Figure 4(a-d) show the antibunching behavior of the
PL lines centered at 744 nm, 749 nm, 756 nm and 764 nm measured with
the HBT interferometer for excitation powers below and above the
optical saturation power $P_\textrm{sat}$. The dip at zero delay
time indicates a single photon emitter. The raw coincidence data were corrected for the
background as described in ref. \cite{Beveratos02b}.

\begin{figure}[htbp]
\centering\includegraphics[width=12cm]{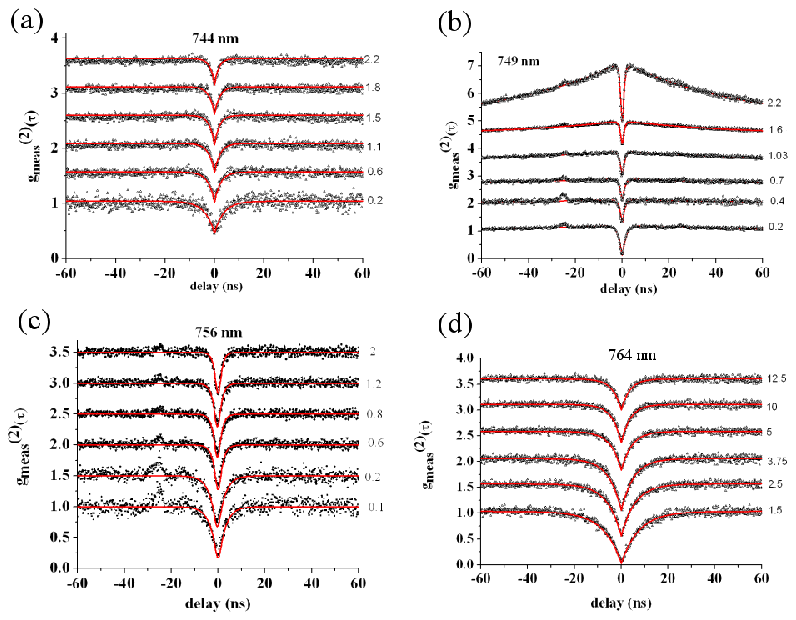} \caption{Background
corrected second-order auto-correlation function
$g^{(2)}_\mathrm{meas}(\tau)$, measured with 154 ps coincidence time
bin for 300 s at different optical powers for the (a) 744 nm line,
(b) 749 nm line the (c) 756 nm line  and the (d) 764 nm line. The
data of the 749 nm line was fit by Eq.(\ref{g2 eq2},\ref{g2 con}),
while the data of the 744 nm, 756 nm and 764 nm lines were fit by
Eq.(\ref{g2 eq1},\ref{g2 con}). The number to the right of the
curves correspond to $P_{\texttt{opt}}/P_{\texttt{sat}}$}
\end{figure}

Due to the observed bunching, the center with a ZPL at 749 nm was fit by Eq.(\ref{g2 eq2}) and Eq.(\ref{g2
con}),which describes three level model. The other three centers, which exhibit two level behavior, were fit by Eq.(\ref{g2 eq1})
and Eq.(\ref{g2 con}). The deviation from zero of the
auto-correlation function is attributed to the jitter of the electronics and detectors and residual polarisation dependent background.

\begin{figure}[htbp]
\centering\includegraphics[width=10cm]{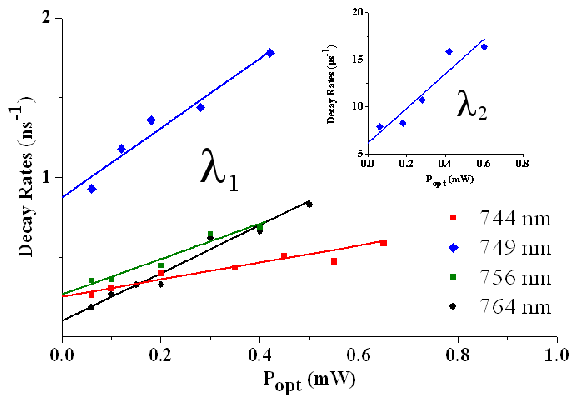} \caption{Estimated
$\lambda_1$ parameters for the centers at 744 nm (red squares), 749
nm (blue diamonds), 756 nm(green triangles) and 764 nm (black
circles) versus the optical power. At the limit of zero optical
power the lifetimes of the centers are respectively
$(r_{21}^0)^{-1}$=3.8 ns, 1.1 ns, 3.7 ns, and 13 ns. The data were
fit with Eq.(\ref{eq4}). The behavior of $\lambda_2$ versus the
optical power of the emitters at 749 nm is also shown inset and fit
with Eq.(\ref{eq4b}); the estimated values $r_{31}^0$ =6.2 MHz (161
ns), and $r_{23}$=0.89 MHz (1.1 $\mu$s), $\alpha=2.5$ mW$^{-1}$,
$\beta=3.1$ mW$^{-1}$.}\label{tau1tau2}
\end{figure}

The $g^{(2)}_{\mathrm{meas}}(0)$ of the 744, 749, 756 and 764 nm centers were 0.44, 0.16, 0.2, and 0.09, respectively. The excited state lifetime of the 744, 756 and 764 nm emitters were 3.8, 3.7 and 13 ns, respectively and were obtained by extrapolating the decay rate $\lambda_1$ to zero optical power, refer to Eq. \ref{eq4}, see Fig. \ref{tau1tau2}. The fit applied to the second order correlation function for the 749 nm emitter contains two exponential decay rates $\lambda_1$ and $\lambda_2$. The extrapolation of these rates to zero optical excitation allows the excited state lifetime and shelving decay rates to be determined. The decay rate $\lambda_2$ as a function of optical power is shown in the inset of Fig. \ref{tau1tau2}. The resulting values of $r_{31}^0$ and $r_{23}$ are 6.2 MHz and 0.89 MHz, respectively and are much smaller than the excited state decay rate, obtained from $\lambda_1$, $r_{21}^0$=880 MHz ($\tau_{21}$ = 1.1 ns).

\begin{figure}[htbp]
\centering\includegraphics[width=10cm]{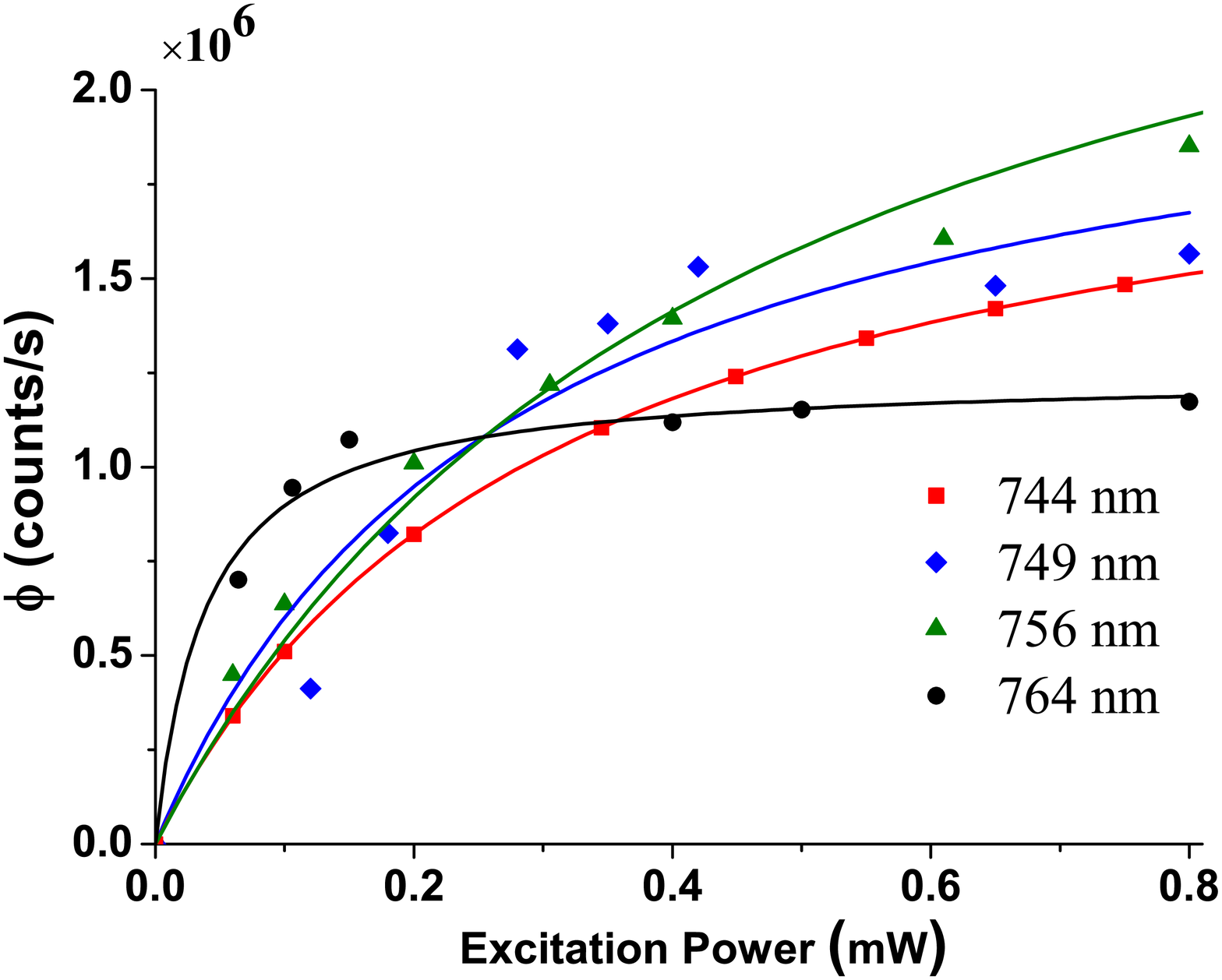} \caption{Measured
saturation curves and fit according to Eq.(\ref{eq3}) with estimated
parameters of optical saturation power, $P_\mathrm{sat}$, and single
photon count rate at saturation, $\phi_{\mathrm{\infty}}$, for the
centers at 744 nm (red squares) with $P_{\mathrm{sat}}$=311 $\mu$W and
$\phi_{\mathrm{\infty}}$=2.1$\times 10^6$ counts/s; at 756 nm (green
triangles) with $P_\mathrm{sat}$=500 $\mu$W and
$\phi_{\mathrm{\infty}}$=3.2$\times 10^5$ counts/s; at 764 nm (black
circles) with $P_\mathrm{sat}$=40 $\mu$W and
$\phi_{\mathrm{\infty}}$=1.3$\times 10^6$ counts/s.  For the center
at 749 nm (blue diamonds) the saturation curve was fit by
Eq.(\ref{QE}) with the estimated values $r_{21}, r_{12}, r_{31},
r_{23}$ and $\eta$. From the fit a quantum efficiency of $\eta_{QE}$=0.24 is
obtained. }\label{satcurve}
\end{figure}

The single photon emission count rate as a function of optical power $P_{\mathrm{opt}}$ is shown for each emitter in Fig.\ref{satcurve}. The measured count rates (corrected for the background) are given by the sum of the counts on the two APDs in the HBT setup. For the 744, 756 and 764 nm emitters, the saturation curves were fit
according to Eq.(\ref{eq3}) yielding saturation count rates $\phi_{\infty}$ of 2.1$\times 10^6$, 3.2$\times 10^6$ and 1.3$\times 10^6$ counts/s,  respectively. The saturation curve of the three
level 749 nm emitter was fit using Eq.(\ref{QE}), with the values of $r_{21}, r_{12}, r_{31}, r_{23}$  obtained from the fit to $g^{(2)}_\mathrm{meas}(\tau)$ as a function of pump power and the value $\eta$ obtained as discussed in the following text. The single fitting parameter $\eta_{QE}$ was then obtained from a least squares fit to the saturation curve.

Figure 7 shows a direct measurement of the lifetime recorded from
the emitters using the pulsed excitation at 20 MHz repetition rate.
The single exponential fit to the fluorescence decay of each emitter, resulted in measured excited state lifetimes of
4.1 ns, 1.4 ns and 14.2 ns for the 744 nm, 749 nm and 764 nm centres, respectively. Upon inspection of the fluorescence decay of the 744 and 749 nm centres there is a fast component of the fluorescence decay which occurs on a time scale less than 0.5 ns, this component is attributed to the background within the emitting crystal. The lifetime of the emitter is this case is determined by fitting the fluorescence decay for times greater than 0.5 ns. The long fluorescence lifetime of the 764 nm centre mitigates this effect and the decay can be described well by the single exponential fit. The measured lifetimes are
in a good agreement with the lifetimes estimated from the CW
measurements for the 744 nm and 749 nm emitters.
The discrepancy with the 764 nm centre can be due to the lack of the CW $g^{(2)}(\tau)$ measurements at
excitation powers well below saturation, which affects the fit to a zero
excitation power.

In the inset of Fig. 7 is an example of the second order correlation function
measurement under pulsed excitation from the 749 nm emitting crystal. The peak at $\tau$=0 is the probability of having two photons
in the same pulse, indicating in this case the presence of only one
photon($g^{(2)}(0)=0.17$). The observed deviation from zero is due
the background luminescence from the diamond crystals, which is not
negligible in this case. A pulsed measurement is mandatory for
quantum optics application since single photons on demand (triggered
single photons) are required. To reduce the unwanted background for practical application, temporal filtering of our
single photon source can be implemented.
\begin{figure}[htbp]
\centering\includegraphics[width=10cm]{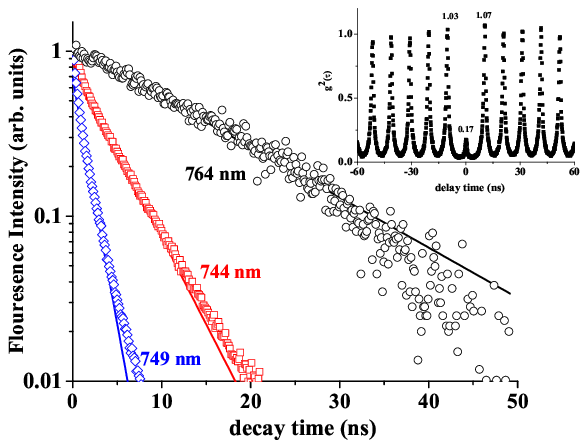} \caption{Direct measurement of the lifetime for the centers at
744 nm (red squares),
749 nm (blue diamonds) and 764 nm (black circles), using a pulsed laser at 20 MHz repetition rate with 200 ps pulse width.
The data were fit with a single exponential. The deduced lifetimes of the centers are respectively
4.1 ns, 1.4 ns and 14.2 ns. Inset: Anti-bunching measurement recorded from a single
 emitter at 749 nm under pulsed laser excitation at 40 MHz and average power of 70 $\mu$W.}
\end{figure}

The total collection efficiency, $\eta$, of our setup can been
 measured directly by exciting a two level emitter in the pulsed regime. The total collection
efficiency can be described by $\eta= \eta_{\mathrm{opt}} \times
\eta_{\mathrm{det}}\times \eta{_\theta}\times \eta_{\mathrm{cr}}$,
where $\eta_{\mathrm{opt}}$ corresponds to the optical
transmittance, $\eta_{\mathrm{det}}$ is the detection efficiency,
$\eta_{\theta}$ corresponds to the geometrical efficiency,
associated to the matching between the collection optics numerical
aperture and the source spatial distribution and $\eta_{\mathrm{cr}}$ accounts for the crystal geometry, that
contributes to the orientation of the excitation dipole of the center with respect to
the excitation electric field and the emission dipole with respect to the collection optics. These are the limiting parameters that reduce the collection efficiency of the source down to typically a few percent.

In the case of a two-level system the total collection efficiency can be measured directly, as each excitation
pulse generates one emitted photon (without being trapped in the metastable state), hence the absence of detected photons is a directly related to $\eta$. In the condition where the laser
excitation energy is above saturation, the laser pulses temporal
separation is longer than the typical detectors dead time (50 ns)
and the source lifetime (14.2 ns), the total collection efficiency can be
directly measured from the total count rate and the laser repetition
rate $\eta= \phi/R_{rep}$. As an example, exciting the 764 nm with a laser repetition rate of 10 MHz,
$\eta=$1.5\% is obtained. Similarly, in the CW regime, $\eta$ is given by $\eta=\phi_{\infty}/(r_{21}^0)$.
By calculating $\eta$ for each two level emitter, an average value of $\eta$=1.3\% is obtained.
The discrepancy in the value of collection efficiency for various
two level emitters may be attributed to a variation in $\eta_{cr}$ due to the unknown dipole orientation of the emitter within the crystal and/or residual polarisation dependent background, that can influence the total saturation count rate $\phi_{\infty}$. The accuracy of $\eta_{cr}$ can be improved by engineering centers with a known dipole orientation and lower fluorescent background.
A detailed analysis of the various parameters of the collection efficiency, which is beyond the
scope of the paper, is however important to exploit fully the two-level system properties for improving
measurement accuracy.
After identifying the collection efficiency $\eta$=1.3\%, the quantum
efficiency $\eta_{QE}$=0.24 of the 749 nm emitter is derived from the fit to the saturation curve with Eq.(9).

 A summary of the photo-physics parameters for each single emitter is shown in Table \ref{table 1}.
\begin{table}[htb]
\centering\caption{Summary of the photo-physics parameters}
 \begin{center}
\begin{tabular}{|c|c|c|c|c|c|c|}
\hline
\raisebox{-1.5ex}[0cm][0cm]{$\lambda$ (nm)} & \raisebox{-1.5ex}[0cm][0cm]{$\eta_{QE}$} & $\phi_{\infty}$  &  $\tau_{21} $ (ns) &  $\tau_{21}$ (ns) & $r_{31}$ & $r_{23}$  \\
& & (counts/s)& (pulsed) & (CW) & (MHz) & (MHz)\\
\hline
744 & 1& 2.1 $10^6$ & 4.1 & 3.8 &&\\
\hline
 756 & 1& 3.2 $10^6$ &  &3.7 & &  \\
\hline
 764 & 1 & 1.3 $10^6$ & 14.2 & 13& &\\
 \hline
749 & 0.24& 2.7 $10^6$ & 1.4 &1.1 & 6.2 &0.89\\
\hline
\end{tabular}
\end{center}\label{table 1}
\end{table}
The ratio between the average off and on
periods of the source is given by the ratio $r_{23}/r_{31}$ which
equals to 0.14 for 749 nm emitter. This value indicates that the
probability of transition to the metastable state is moderate, as
shown by the slight bunching of the $g^{(2)}(\tau)$ function (Fig.
4b). Previous determinations associated to NE8 complex in bulk
natural diamond IIa \cite{Gaebel04,Wu06}, and CVD nanodiamond
\cite{Wu07} showed $r_{23}/r_{31}$ of 2.8, 1.6 and 0.8 respectively,
justifying much lower saturation rates of NE8 with respect to the 749 nm center. In addition, the center possess a
shorter lifetime than NE8, which also contributes to the higher emission rates
despite the moderately low quantum efficiency.

In general two level emitters possessed a ZPL with a FWHM of $\sim$ 10 nm, whilst the three level emitters exhibited a FWHM of $\sim$ 4 nm. The variation of the lifetime (4 ns to 14 ns) of the two level system emitters can be
attributed to nanocrystals geometry experiencing local modification of electric field and/or differences in the atomic structure.

The single emitting centres in the range 740-770 nm are believed to arise from chromium atoms within the diamond lattice. The charge state of the chromium complex and number of atoms involved is unknown. The explanation of what leads to two or three level behavior of a particular crystal is still under investigation.
One may assume that the strain within the crystal may have a significant effect.
Indeed, it was shown for NV centers that strain can modify the transition
from being spin conserving to a $\Lambda$ system geometry\cite{Santori06}.
Similarly, the strain can modify the electronic transition and enhance/suppress the relaxation to a third metastable state.

\section{Conclusion}

A full optical characterization of the photophysics of
novel optical centers attributed to chromium related defects in
sub-micron CVD grown diamond crystals was performed.  Novel two-level systems with ZPLs at 744 and 764nm, were identified and characterized. The direct measurement of the total collection efficiency from the two level systems enabled a more accurate determination of the quantum efficiency of the three level single emitter at 749 nm ($\eta_{QE}$=0.24).

The spectral width of the two level emitters was $\sim$ 10 nm, whilst the three level emitter exhibited a FWHM of $\sim$ 4 nm. This coupled with the
short excited state lifetimes (1-14 ns) and the high brightness (up to MHz
regime), brings these diamond centers much closer to an ``ideal
single photon source on demand''. The precise engineering of two-level single photon emitters may lead to such sources being implemented as a single photon standard.

\section{Acknowledgment}
This work was supported by the Australian Research Council, The
International Science Linkages Program of the Australian Department
of Innovation, Industry, Science and Research (Project No. CG110039)
and by the European Union Sixth Framework Program under the EQUIND
IST-034368. ADG is the recipient of an Australian Research Council Queen Elizabeth II Fellowship (project No. DP0880466).

\end{document}